\documentclass{PoS}

\usepackage[T1]{fontenc}
\usepackage[latin9]{inputenc}
\usepackage{textcomp}
\usepackage{amsbsy}
\usepackage{graphicx}

\title{Dynamical Structure of Baryons}

\ShortTitle{Dynamical Structure of Baryons}

\author{\speaker{A. Aleksejevs}
\\
        Grenfell Campus of Memorial University\\
        E-mail: \email{aaleksejevs@grenfell.mun.ca}}

\author{S. Barkanova\\
        Acadia University\\
        E-mail: \email{svetlana.barkanova@acadiau.ca}}

\abstract{Compton scattering offers a unique opportunity to study the dynamical
structure of hadrons over a wide kinematic range, with polarizabilities
characterizing the hadron\textquoteright{}s active internal degrees
of freedom. We present calculations and detailed analysis of electric
and magnetic and the spin-dependent dynamical polarizabilities for
the lowest in mass SU(3) octet of baryons. These extensive calculations
are made possible by the recent implementation of semi-automatized
calculations in chiral perturbation theory which allows evaluating
polarizabilities from Compton scattering up to next-to-the-leading
order. The dependencies for the range of photon energies covering
the majority of the meson photoproduction channels are analyzed.}

\FullConference{36th International Conference on High Energy Physics,\\
		July 4-11, 2012\\
		Melbourne, Australia}

\begin{document}

\section{Introduction}

One of the major goals of the low-energy QCD is the investigation
of the baryon response to the external electromagnetic field via multipole
excitation mechanism. Structure parameters which describe that response
are electric, magnetic and spin-dependent polarizabilities. In other
words, the polarizabilities are related to the deformability and stiffness
of the baryon. Precise determination of nucleon polarizabilities still
requires substantial effort from both theory and experiment. For hyperons,
the polarizabilities are yet to be measured. In this work, we study
the polarizabilities of baryons using the Compton scattering, which
is a straightforward process from both theoretical and experimental
points of view. In general, the polarizabilities in a very low energy
region of the Compton scattering are treated as static (with very
little or no dependence on the photon energy), but it can be assumed
that at higher energies, and especially near meson production threshold,
this static behavior will break and the polarizabilities will become
dynamic. The main goal of this work is a study of dependence of the
electric, magnetic and spin-dependent polarizabilities on the photon
energy. We use the relativistic chiral perturbation theory (ChPTh)
while applying the multipole expansion approach for the Compton structure
functions. The various versions of ChPTh predict a rather broad spectrum of 
values for polarizabilities, but so far, it is only theory available in the 
regime of the non-perturbative QCD and have been employed here using our computational
 hadronic model (CHM \cite{CHM}). CHM gives us a possibility
to avoid the low-energy approximation in the Compton structure functions
and retain all the possible degrees of freedom arising from SU(3)
chiral Lagrangian. We provide a short description of the formalism
used in this work in the section \textquotedblleft{}Formalism\textquotedblright{}.
Analysis of the dynamical behavior of polarizabilities along with
their static values is presented in the \textquotedblleft{}Results\textquotedblright{}.

\section{Formalism}

In the presence of external electromagnetic field, induced electric
and magnetic dipole moments of the baryon generate 
 effective Hamiltonian $H_{eff}=-\frac{1}{2}4\mathbf{\pi}\alpha{\bf E}^{2}-\frac{1}{2}4\pi\beta{\bf H}^{2}$.
Here, proportionality constants $\alpha$ and $\beta$ are called
electric and magnetic polarizabilities, respectively. 
Although the polarizability values are quite small ($10^{-4}\,(fm^{3})$),
they were successfully measured by several experimental groups using the Compton scattering
 and  employing the dispersion sum rules analysis to extract the polarizabilities from the cross
 section data. The current PDG \cite{PDG} averaged experimental
values for the electric and magnetic polarizabilities for proton and
neutron are:
\begin{eqnarray*}
\alpha_{p}=(12.0\pm0.6)10^{-4}(fm^{3}); &  & \beta_{p}=(1.9\pm0.5)10^{-4}(fm^{3});\\
\alpha_{n}=(11.6\pm1.5)10^{-4}(fm^{3}); &  & \beta_{n}=(3.7\pm2.0)10^{-4}(fm^{3}).
\end{eqnarray*}
For proton and neutron $\alpha$ and $\beta$ values are approximately
the same, and the positive value of the magnetic polarizability
points to the paramagnetic nature of the nucleon. 

If the baryon is placed in the time-varying electromagnetic
field, another set of dipole moments is induced and the following
effective Hamiltonianl describes that type of interaction:
\begin{eqnarray}
 & H_{eff}^{spin}=-\frac{1}{2}2\pi\gamma_{E1E1}\boldsymbol{\sigma}\cdot(\mathbf{E}\times\mathbf{\dot{E}})-
\frac{1}{2}2\pi\gamma_{M1M1}\boldsymbol{\sigma}\cdot(\mathbf{B}\times{\bf \dot{B}})-4\pi\gamma_{M1E2}\sigma_{i}B_{j}E_{ij}-4\pi\gamma_{E1M2}\sigma_{i}E_{j}B_{ij}\nonumber \\
\label{eq:2}\\
 & T_{ij}=\frac{1}{2}(\partial_{i}T_{j}+\partial_{j}T_{i});\,\,\,\mathbf{T}=\{\mathbf{E},\,\mathbf{B}\}.\nonumber 
\end{eqnarray}
The coefficients of the proportionality in the Eq.(\ref{eq:2}), $\gamma_{E1E1},\,\gamma_{M1M1},\,\gamma_{M1E2}$
and $\gamma_{E1M2}$ are called spin-dependent polarizabilities and
correspond to the dipole-dipole and dipole-quadrupole electric/magnetic
transitions. It is quite difficult to measure these spin-dependent
polarizabilities separately, but for the specific kinematics of the forward/backward
scattering these structure parameters can be combined into so-called
forward $\gamma_{0}$ and backward $\gamma_{\pi}$ polarizabilities,
$\gamma_{0}=-\gamma_{E1E1}-\gamma_{M1M1}-\gamma_{M1E2}-\gamma_{E1M2}$ and  
$\gamma_{\pi}=-\gamma_{E1E1}+\gamma_{M1M1}+\gamma_{M1E2}-\gamma_{E1M2}$, 
and can be accessed by the experiment.
In order to evaluate the polarizabilities theoretically, one can use the Compton
scattering and relate the amplitude to the set of the Compton structure
functions $R_{i}$ \cite{Babusci} in the following way:
\begin{eqnarray}
 & \frac{1}{8\pi W}M(\gamma B\rightarrow\gamma'B)= & R_{1}(\boldsymbol{\epsilon}{}^{\prime*}\cdot\boldsymbol{\epsilon})+R_{2}({\bf s}{}^{\prime*}\cdot{\bf s})+iR_{3}\boldsymbol{\sigma}\cdot(\boldsymbol{\epsilon}{}^{\prime*}\times\boldsymbol{\epsilon})+iR_{4}\boldsymbol{\sigma}\cdot({\bf s}{}^{\prime*}\times{\bf s})+\nonumber \\
\label{eq:4} \\
 &  & iR_{5}((\boldsymbol{\sigma}\cdot\hat{{\bf k}})({\bf s^{\prime*}}\cdot\boldsymbol{\epsilon})-(\boldsymbol{\sigma}\cdot\hat{{\bf k}}^{\prime})({\bf s}\cdot\boldsymbol{\epsilon^{\prime*}}))+iR_{6}((\boldsymbol{\sigma}\cdot\hat{{\bf k}}^{\prime})({\bf {\bf s^{\prime*}}}\cdot\boldsymbol{\epsilon})-(\boldsymbol{\sigma}\cdot\hat{{\bf k}})({\bf s}\cdot\boldsymbol{\epsilon^{\prime*}}))\nonumber
\end{eqnarray}
Here, $W=\omega+\sqrt{\omega^{2}+m_{B}^{2}}$ is the center of mass
energy and $\omega$ is the energy of the incoming photon. The unit magnetic
vector (${\bf s}=(\hat{{\bf k}}\times\boldsymbol{\epsilon}$)), polarization
vector ($\boldsymbol{\epsilon}$) and unit momentum of the photon
(${\displaystyle \hat{{\bf k}}=\frac{{\bf k}}{k}}$) are denoted by
the prime for the case of the outgoing photon. Although the choice
of the basis for the invariant Compton amplitude is not unique \cite{Babusci}, the basis in the Eq.(\ref{eq:4})
is the most convenient for evaluating polarizabilities.
Here, in Eq.(\ref{eq:4}), the structure functions $R_{i}$ are directly
related to the electric, magnetic and spin-dependent polarizabilities
in the multipole expansion. 
If we keep only dipole-dipole and dipole-quadrupole transitions in
the multipole expansion of the Compton structure functions \cite{multi-1},
we have rather simple connecting formulas to the polarizabilities
of the baryon: 
\begin{eqnarray}
 & {\displaystyle R_{1}^{NB}=\omega^{2}\alpha_{E1};\ \ R_{2}^{NB}=\omega^{2}\beta_{M1};\ \ R_{3}^{NB}=\omega^{3}(-\gamma_{E1E1}+\gamma_{E1M2});}\nonumber \\
\label{eq:6}\\
 & {\displaystyle R_{4}^{NB}=\omega^{3}(-\gamma_{M1M1}+\gamma_{M1E2});\ \ R_{5}^{NB}=-\omega^{3}\gamma_{M1E2};\ \ R_{5}^{NB}=-\omega^{3}\gamma_{E1M2}.}\nonumber
\end{eqnarray}
Although the polarizabilities used in Eq.(\ref{eq:6}) are defined as
constants, it is essential to treat them as energy-dependent quantities
\cite{Griesshammer}. The reason behind this extension to the dynamical
(energy-dependent) polarizabilities is dictated by the fact that the
Compton scattering experiments were performed with 50 to 800 MeV photons
and hence required additional theoretical information to extrapolate
the results to zero-energy parameters. It is also well known
that the polarizabilities can become energy-dependent due to the internal
relaxation mechanisms, resonances, and particle production thresholds. Accordingly, we keep all orders in $\omega$
for the Compton structure functions (for static polarizabilities
we keep only order up to $\mathcal{O}(\omega^{2})$ for $R_{1,2}$
and up to $\mathcal{O}(\omega^{3})$ for $R_{3,4,5,6}$), and hence
determine the energy-dependent polarizabilities. A connection
between dynamic and static polarizabilities can be achieved
by taking a limit to zero photon energy. The Compton
structure functions up to one-loop order are calculated using
CHM \cite{CHM} based on the relativistic chiral perturbation
theory. In addition, the structure-dependent pole contribution to the
nucleon polarizabilities is taken into account in the form of
the nucleon $\Delta$-resonance excitation. 

\section{Results}

\begin{figure}
\begin{centering}
\begin{tabular}{cc}
\includegraphics[scale=0.2]{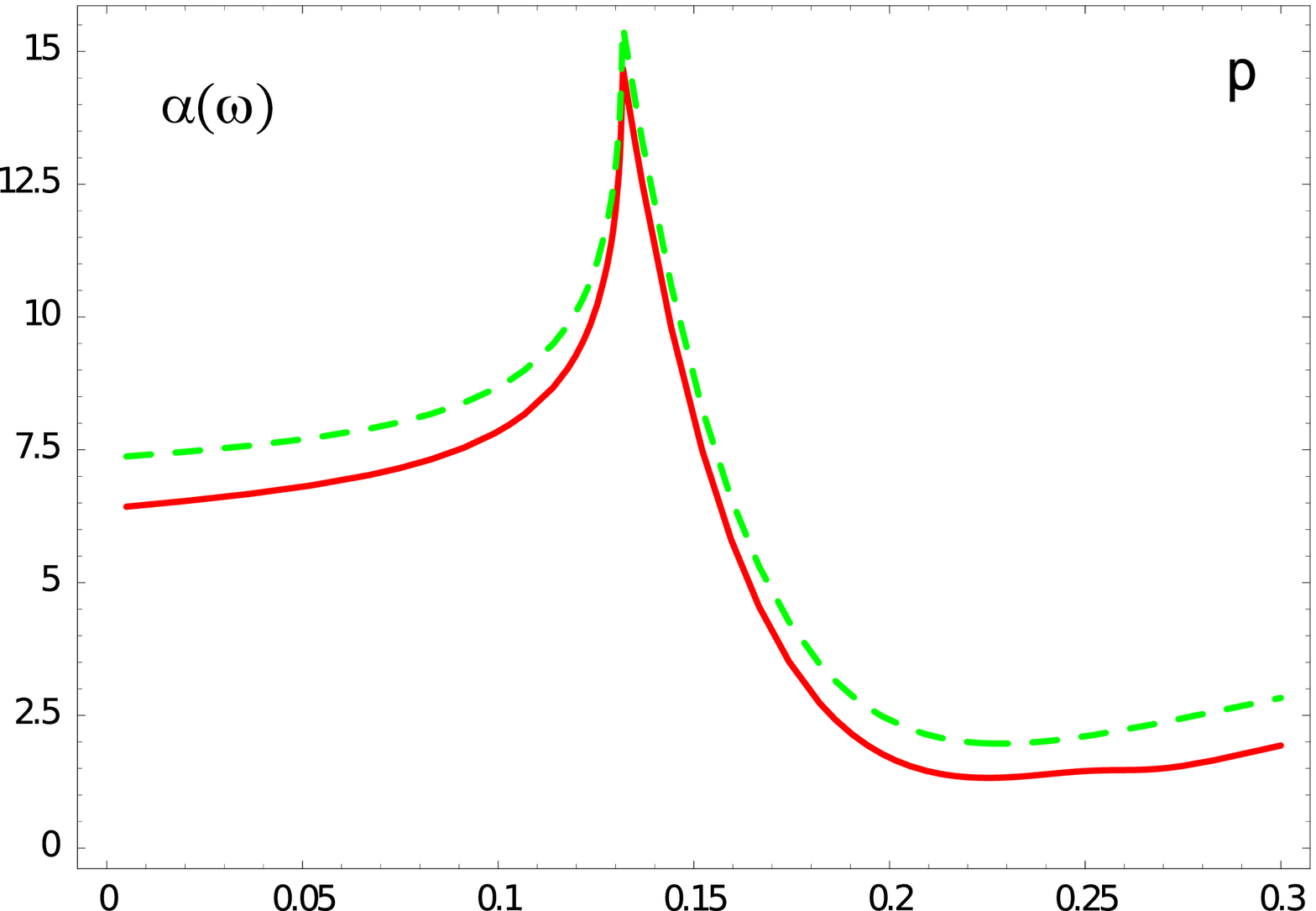} & \includegraphics[scale=0.2]{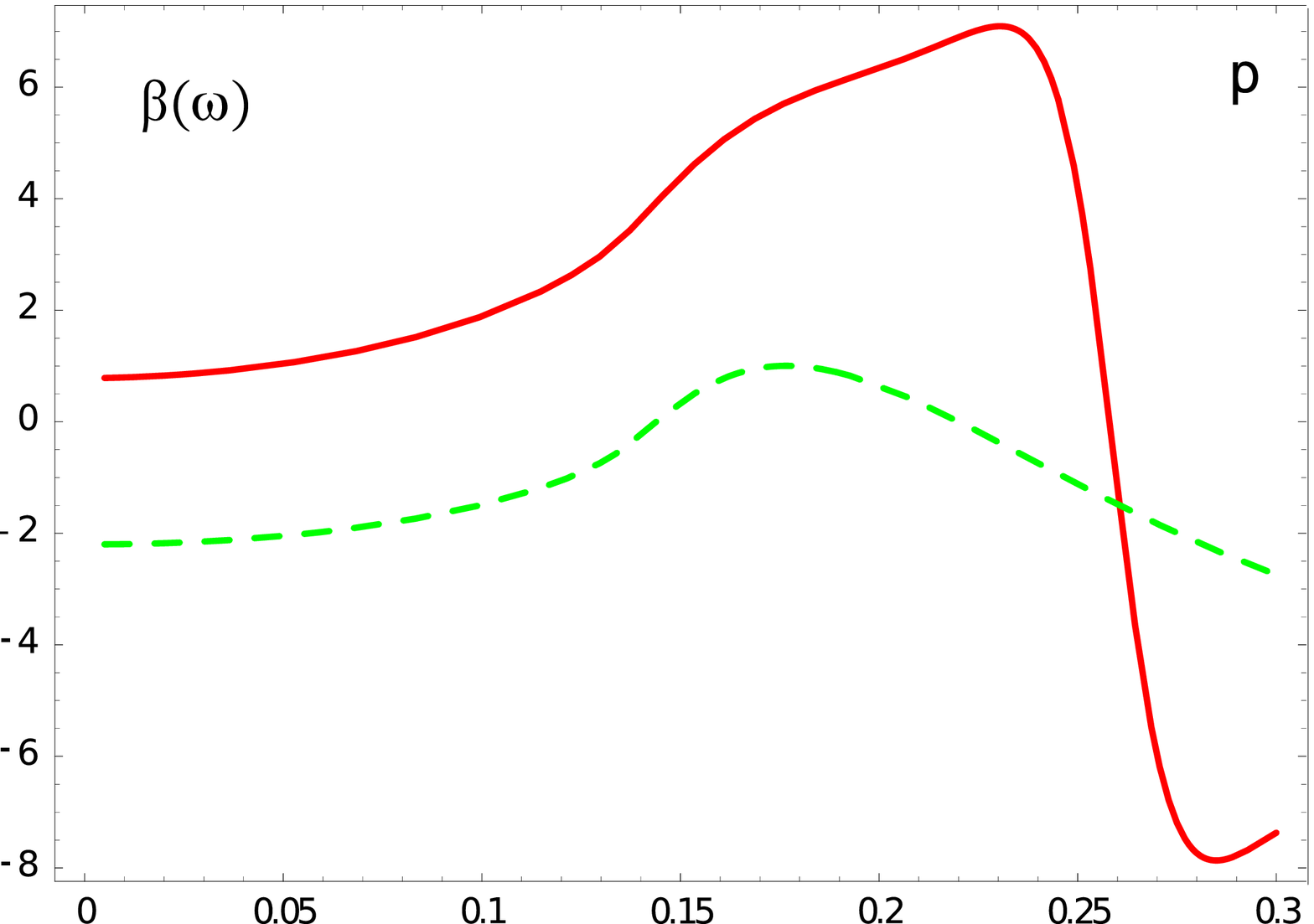}\tabularnewline
\includegraphics[scale=0.2]{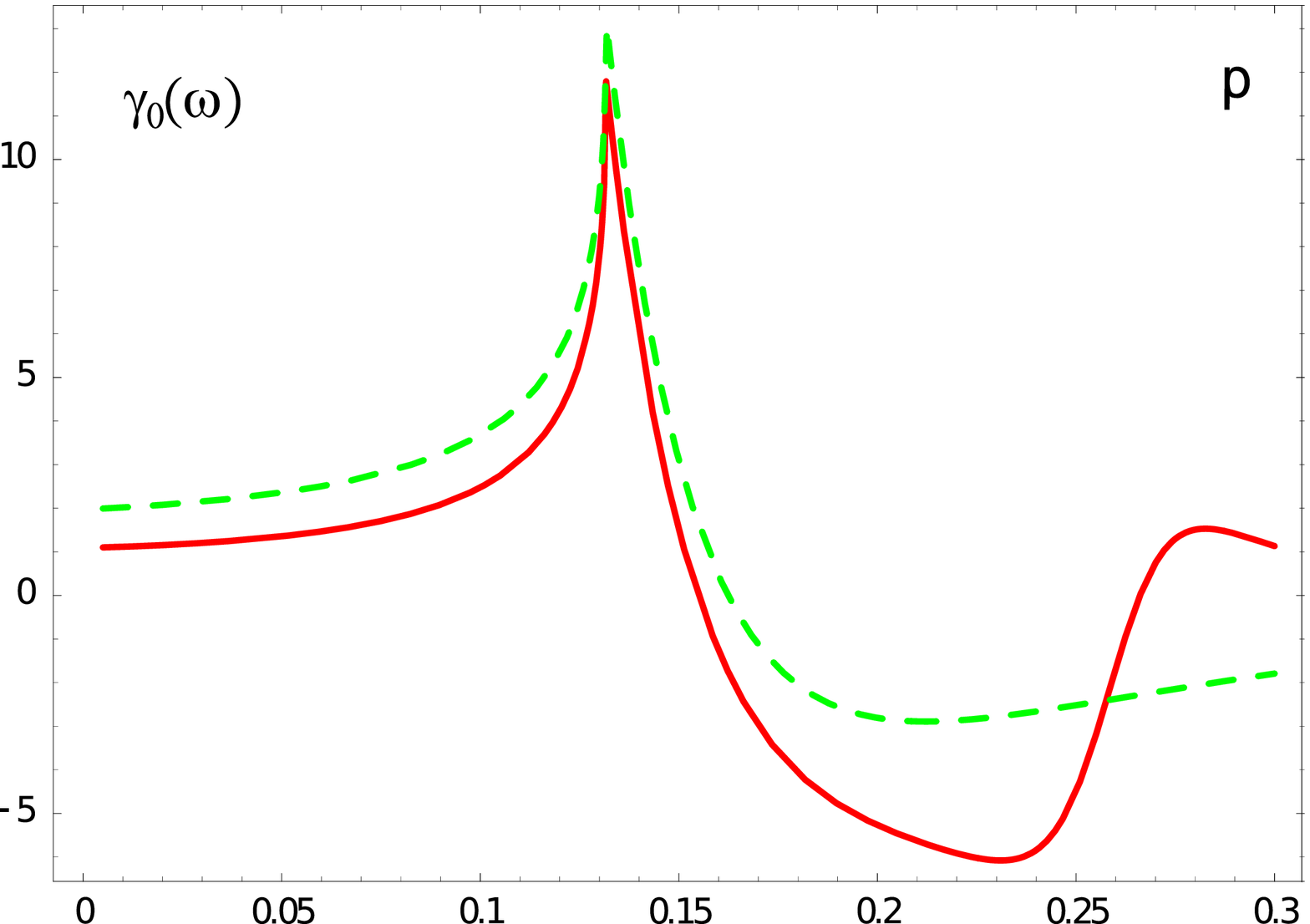} & ~\includegraphics[scale=0.2]{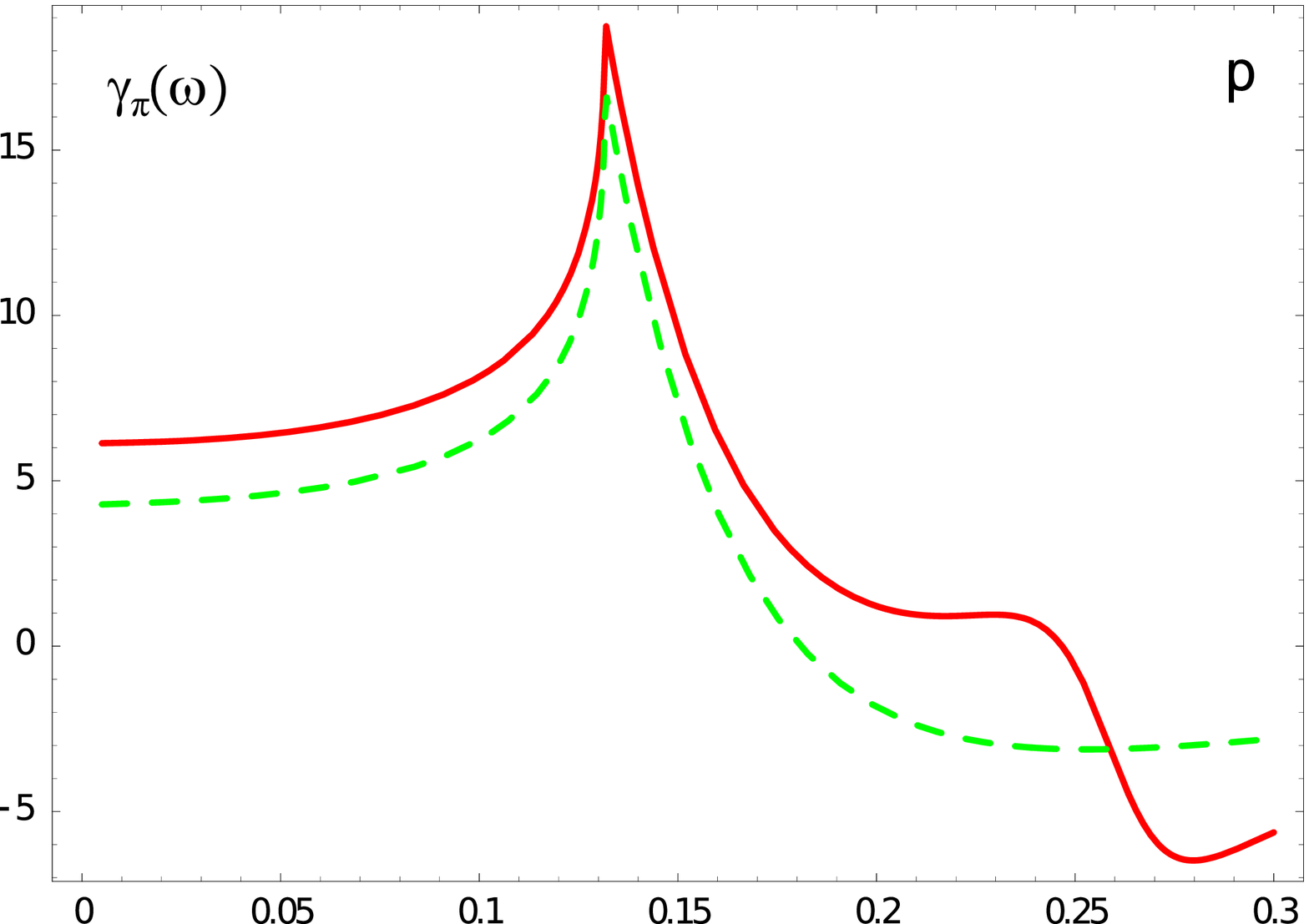}\tabularnewline
\end{tabular}\\

\par\end{centering}

\centering{}\caption{Dependencies of the proton electric, magnetic and spin-dependent polarizabilities
on photon energy $\omega$(GeV) in the center-of-mass reference frame.
The first row (from left) corresponds to the electric and magnetic
polarizabilities in $10^{-4}\,(fm^{3}).$ The second row are spin-dependent forward $\gamma_{0}$ and backward $\gamma_{\pi}$
polarizabilities in $10^{-4}\,(fm^{4}).$ Green-dashed curve corresponds
to the meson-nucleon loops contribution only and solid-red curve
is the result with $\Delta$ resonance pole contribution added.}
\label{ff1}
\end{figure}

The polarizabilities calculated for the proton with the photon energies
up to 300 MeV are shown on Fig.\ref{ff1}. It is evident that
below 50 MeV they have very small energy dependence. For the neutron,
the energy dependencies of the dynamical polarizabilities have similar
behavior except the values are bigger on absolute scale. Here
we will only provide a description for the proton dynamical polarizabilities. 

\begin{figure}
\begin{centering}
\begin{tabular}{ccc}
\includegraphics[scale=0.2]{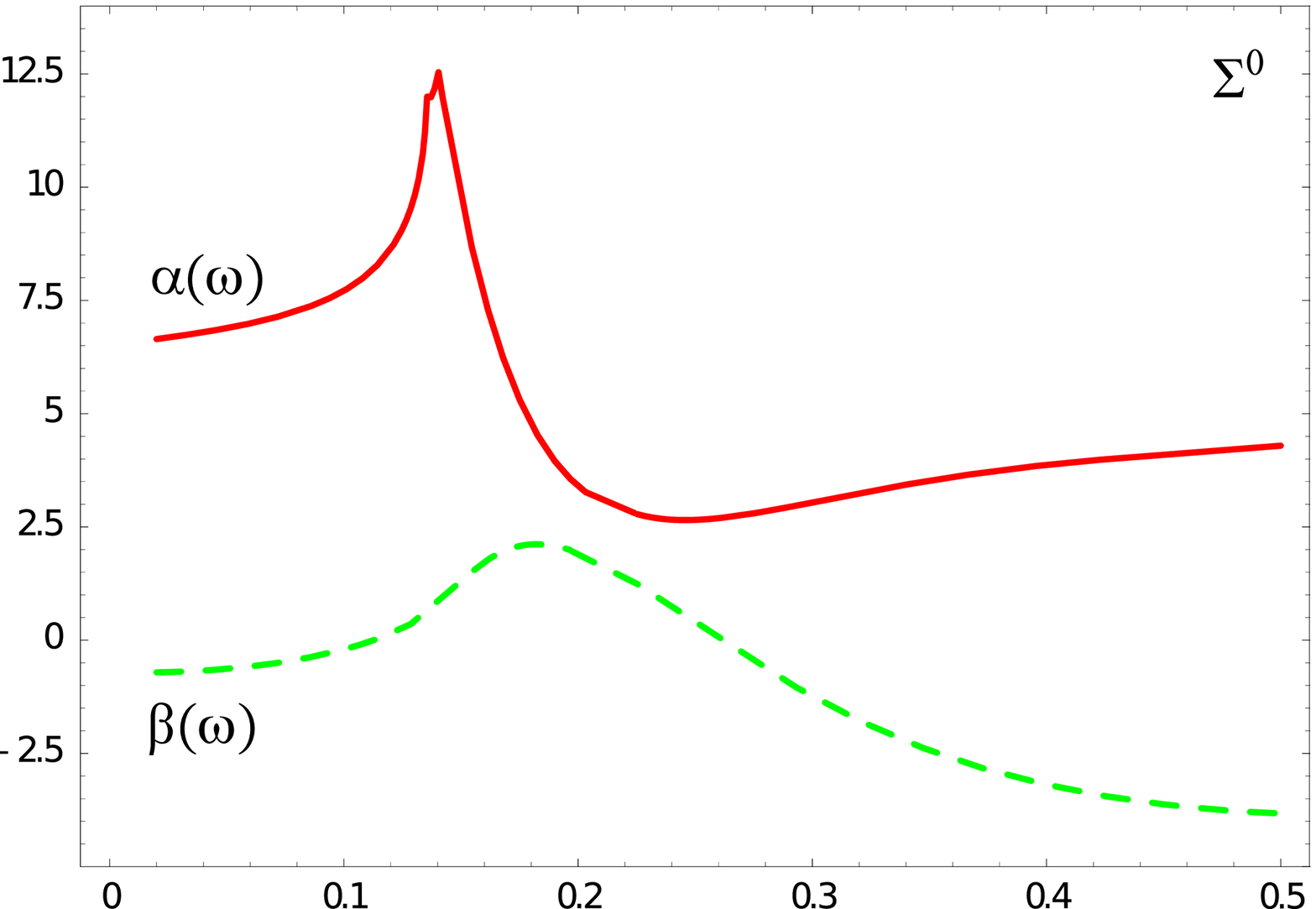} & \includegraphics[scale=0.2]{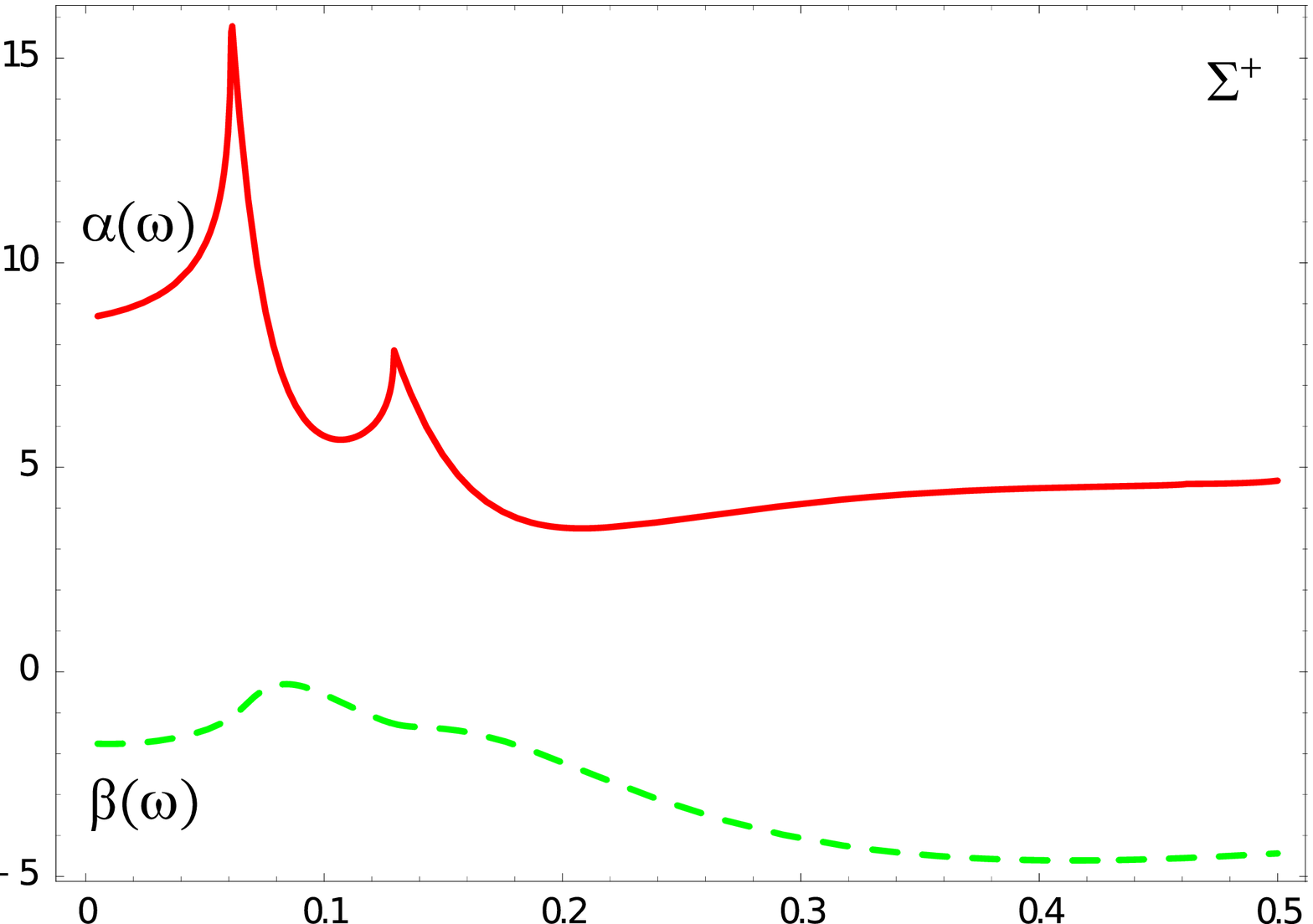} & \includegraphics[scale=0.2]{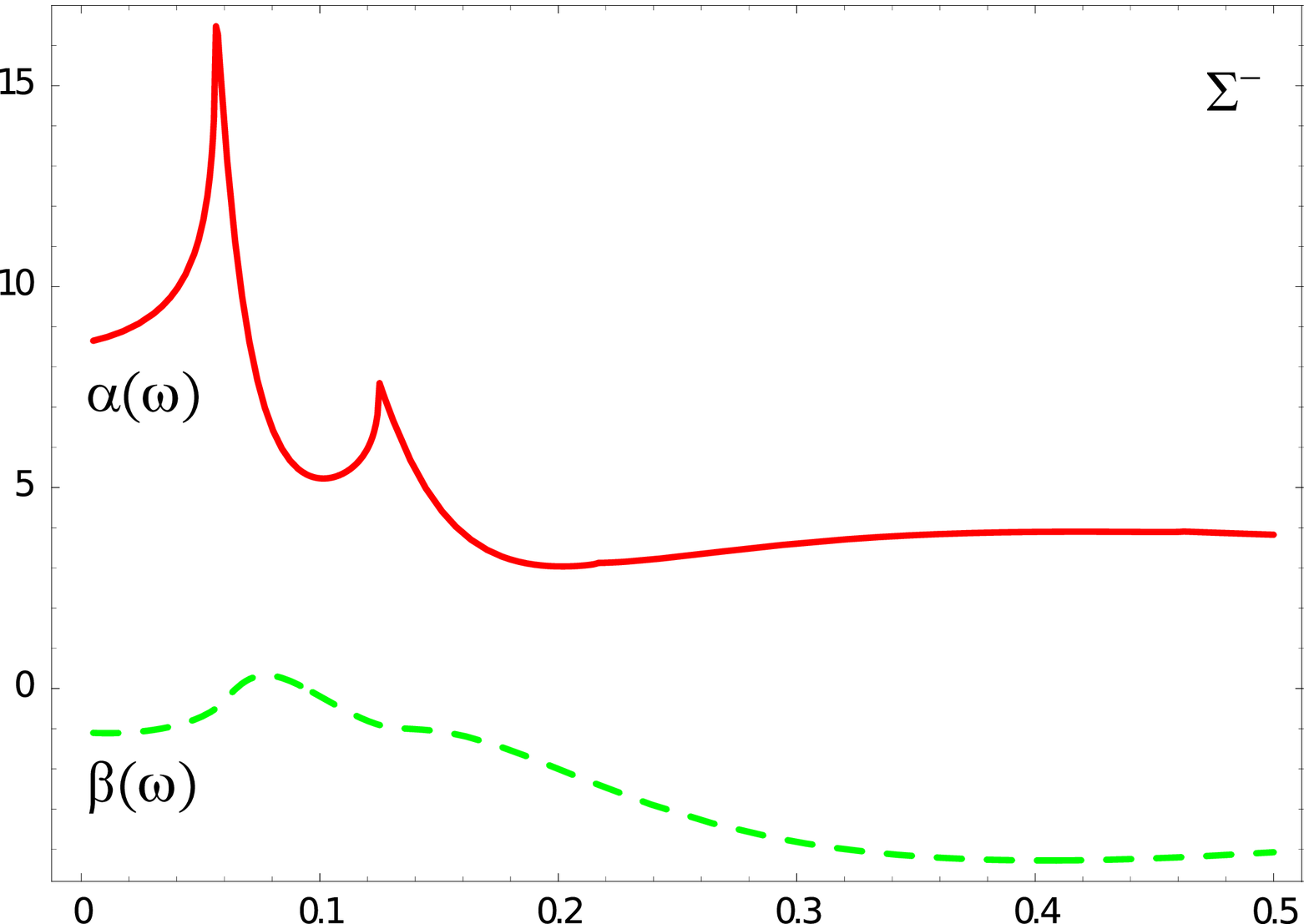}\tabularnewline
\includegraphics[scale=0.2]{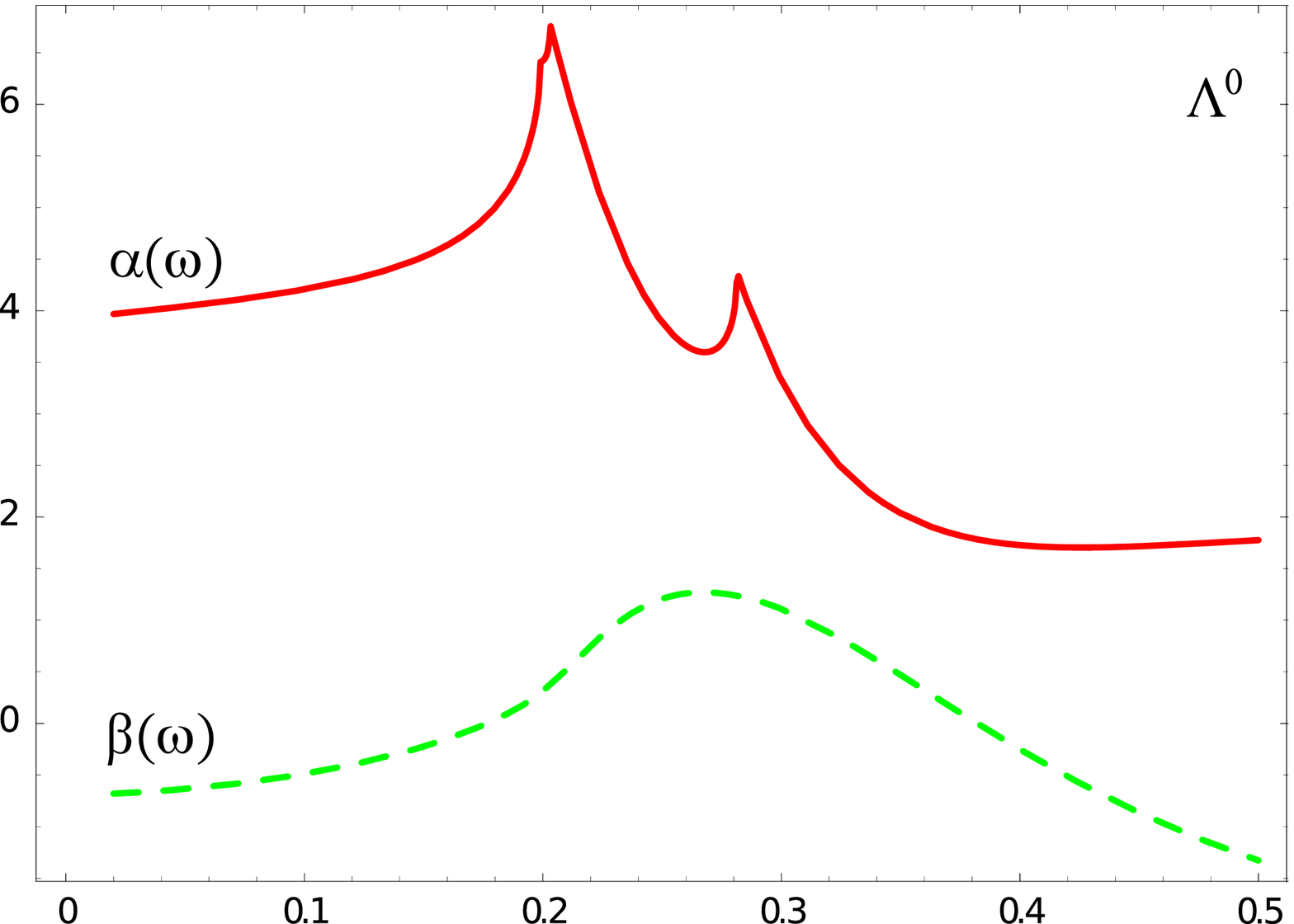} & \includegraphics[scale=0.2]{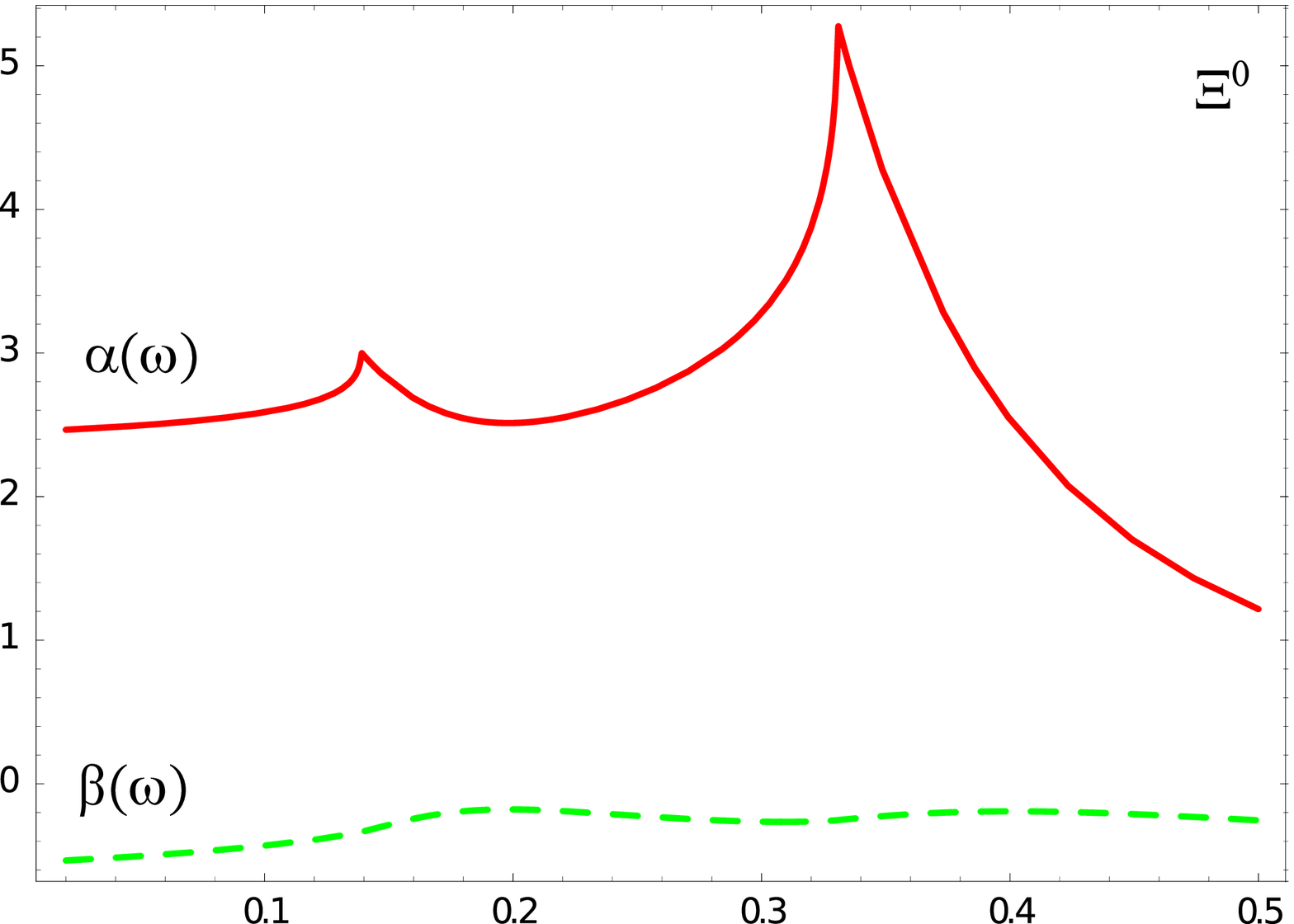} & \includegraphics[scale=0.2]{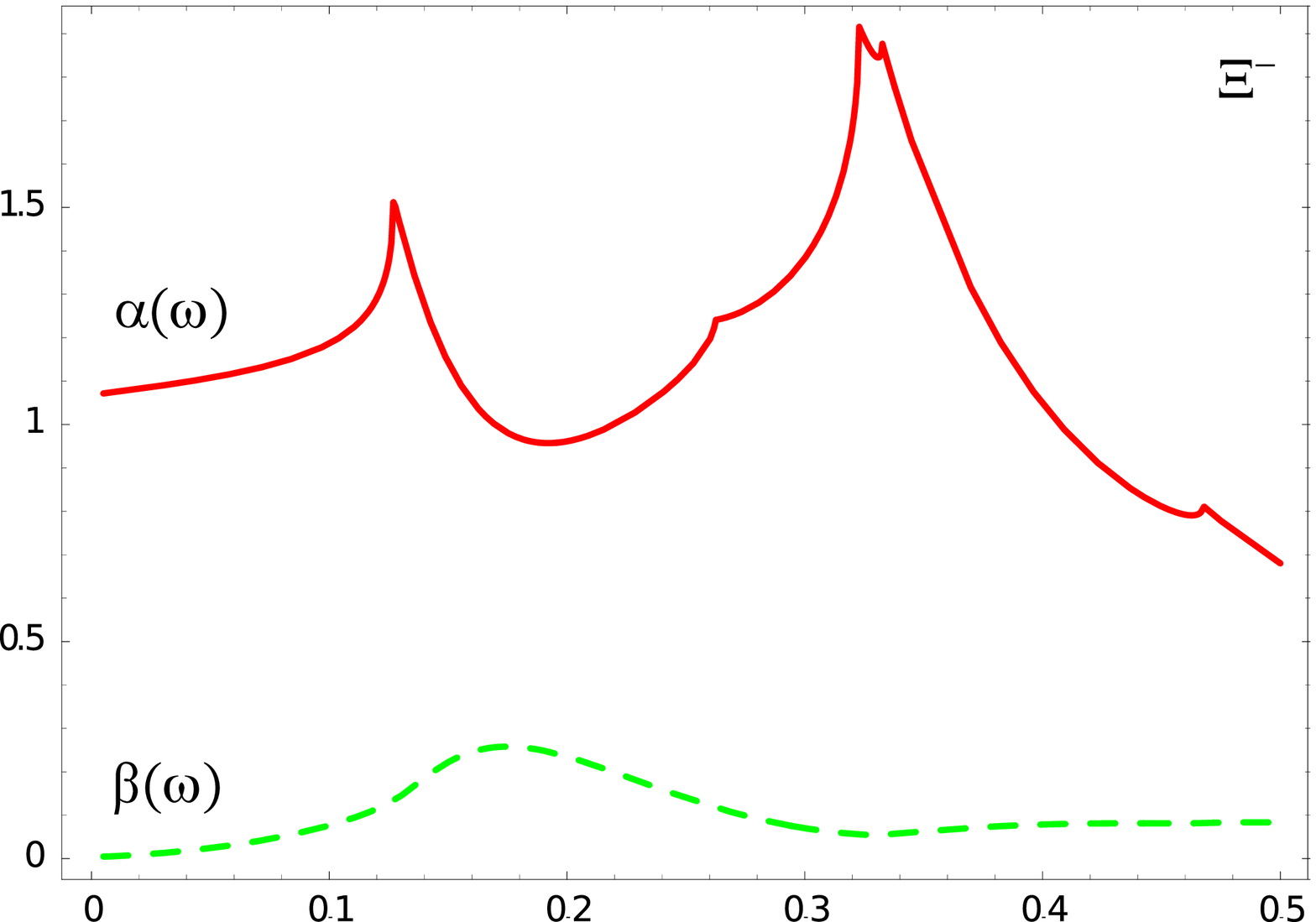}\tabularnewline
\end{tabular}\\

\par\end{centering}

\caption{Electric and magnetic dynamical polarizabilities of hyperons in units
of $10^{-4}\,(fm^{3})$ as a function of the photon energy $\omega$(GeV).
Here, solid red line represents the electric polarizability and dashed green line is
the magnetic polarizability.}

\label{ff2}
\end{figure}

The electric polarizability of the proton has very strong, resonance-type
dependence near the pion production threshold. The $\Delta$-pole
contribution has a small effect while consistently reducing $\alpha_{p}(\omega)$
values for all the energies. Of course, to make final predictions
of the ChPTh of the values of polarizabilities, it is required to
add contribution from the resonances in the Compton scattering loops.
Hence, in order to compare our results with experimental values, we
have used resonance loops results borrowed from the small scale expansion
(SSE) approach \cite{SSE}. If no $\Delta$-pole contribution is
added, the magnetic polarizability in Fig.\ref{ff1} stays negative (diamagnetic)
for almost all the energies. The $\Delta$-pole contribution is very
large and shifts $\beta_{p}(\omega)$ from negative to positive (paramagnetic)
values for energies up to 250 MeV. This behavior is quite natural,
since the pion loop calculations reflect magnetic polarizability coming
from the virtual diamagnetic pion cloud and the $\Delta$ resonance
contribution to $\beta_{p}(\omega)$ is driven by the strong paramagnetic
core of the nucleon. The spin-dependent polarizabilities, $\gamma_{0}$
and $\gamma_{\pi}$, have strong dependence near the pion production
threshold and the $\Delta$- pole contribution is evident near the $\Delta$
production threshold. If we take contributions of order $\mathcal{O}(p^{3})$
in ChPTh power counting, we get an excellent agreement with \cite{LenPas}.
Our result for the proton polarizabilities up to the one-loop order, plus
including $\Delta$- pole and SSE contribution is  the following (in units
of $10^{-4}(fm^{3})$):
\begin{eqnarray*}
 & \alpha_{p}=(7.38\,(\pi-\mbox{loop})-0.95\,(\Delta-\mbox{pole})+4.2\,(\mbox{SSE}))=10.63;\\
 & \beta_{p}=(-2.20\,(\pi-\mbox{loop})+3.0\,(\Delta-\mbox{pole})+0.7\,(\mbox{SSE}))=1.49.
\end{eqnarray*}
In the following table, we list the results for the spin-dependent static polarizabilities:
\begin{center}
\begin{tabular}{|c|c|c|c|c|c|c|}
\hline 
$10^{-4}\,(fm^{4})$ & $\mathcal{O}(p^{3})$\cite{OP3} & $\mathcal{O}(p^{4})$\cite{OP4} & $\mathcal{O}(\epsilon^{3})$\cite{SSE2} & HYP. Dr. \cite{HYPDR} & This work & Exp.\tabularnewline
\hline 
\hline 
$\gamma_{0(p)}$ & 4.6 & -3.9 & 2.0 & -1.1 & 1.1 & -0.90 \textpm{}0.08 \textpm{}0.11\tabularnewline
\hline 
$\gamma_{\pi(p)}$ & 4.6 & 6.3 & 6.8 & 7.8 & 6.1 & 8.0 \textpm{}1.8\tabularnewline
\hline 
\end{tabular}
\par\end{center}
The listed results have a broad spectrum of values, so clearly more work is needed in this area. Our values in this table do not include the
$\Delta$ resonance in the loops, but if we follow the trend of the
$\Delta$- pole contribution into $\gamma_{0}$ and $\gamma_{\pi}$,
we can see that inclusion of resonance in the loops for the Compton scattering
will bring our results closer to the experimental values. 
The static electric and magnetic polarizabilities for hyperons have been
calculated first in \cite{Meissner} and just recently calculations
have been completed for the spin-dependent static polarizabilities
in \cite{Kumar}. Both groups were using heavy baryon chiral perturbation
theory. The dynamical electric and magnetic polarizabilities for hyperons
first have been calculated in \cite{AB}. In Fig.\ref{ff2} we provide
updated results for dynamical electric and magnetic polarizabilities
for hyperons using basis from Eq.(\ref{eq:4}) in the Compton scattering
amplitude. 

\begin{figure}
\begin{centering}
\begin{tabular}{ccc}
\includegraphics[scale=0.2]{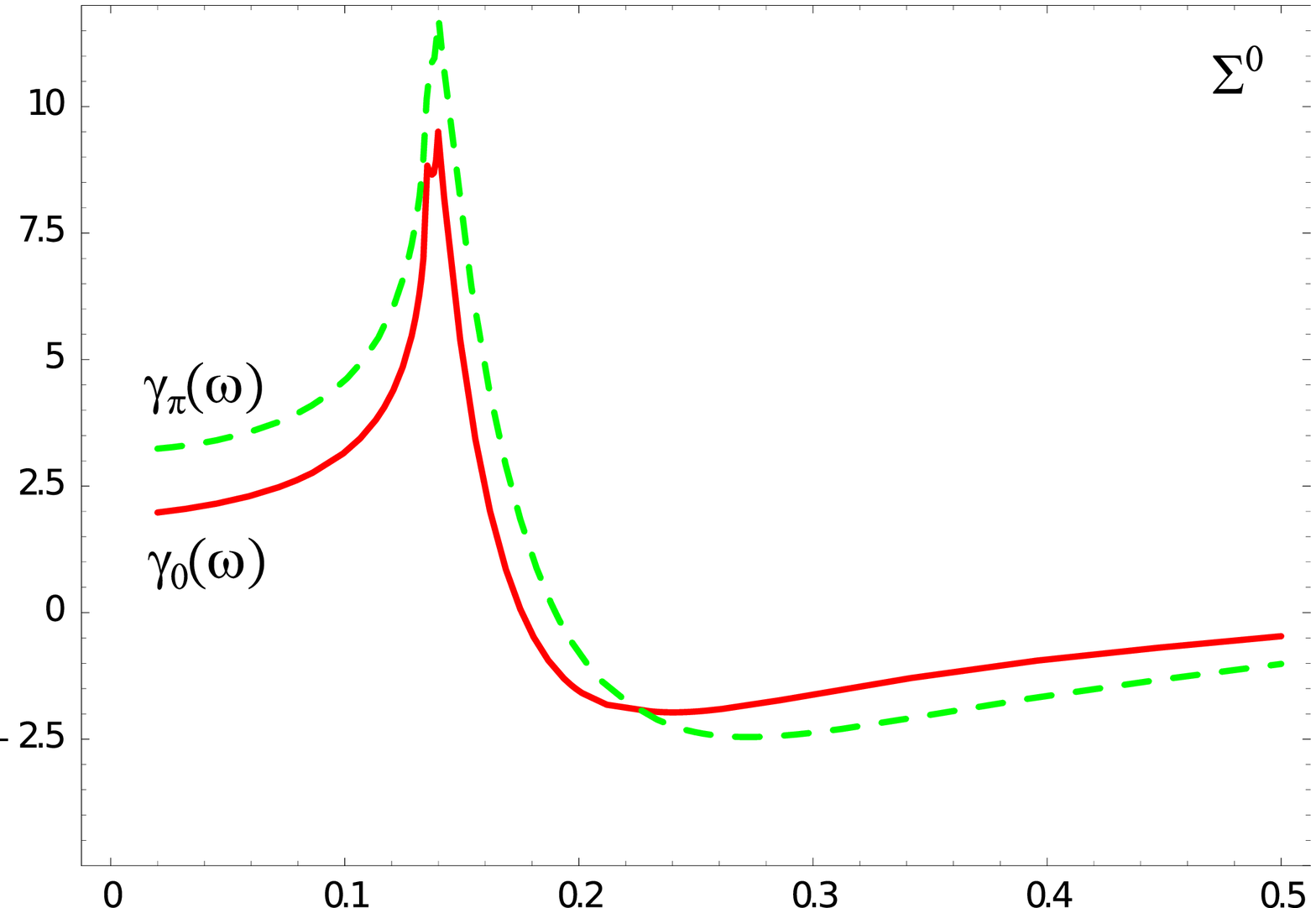} & \includegraphics[scale=0.2]{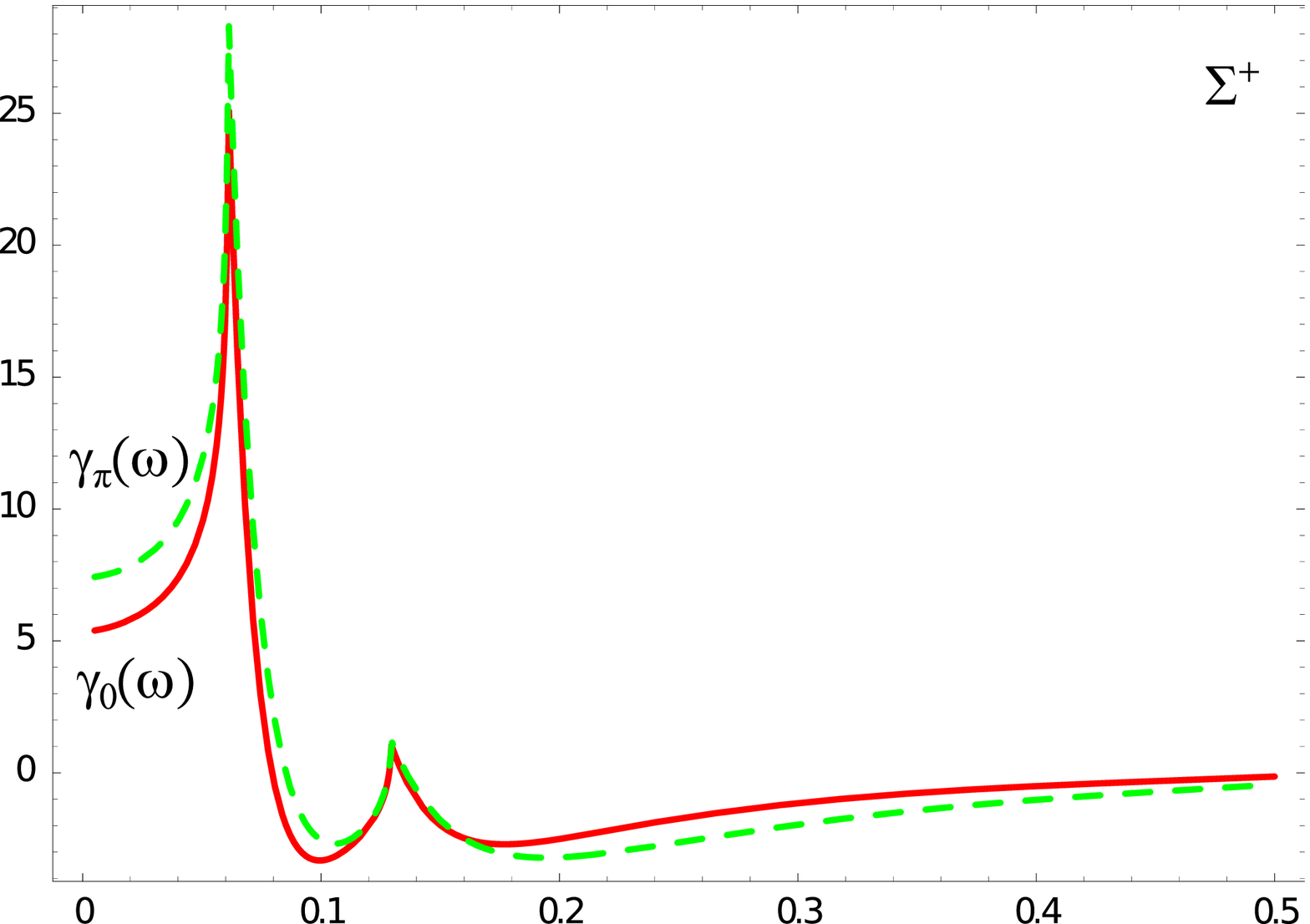} & \includegraphics[scale=0.2]{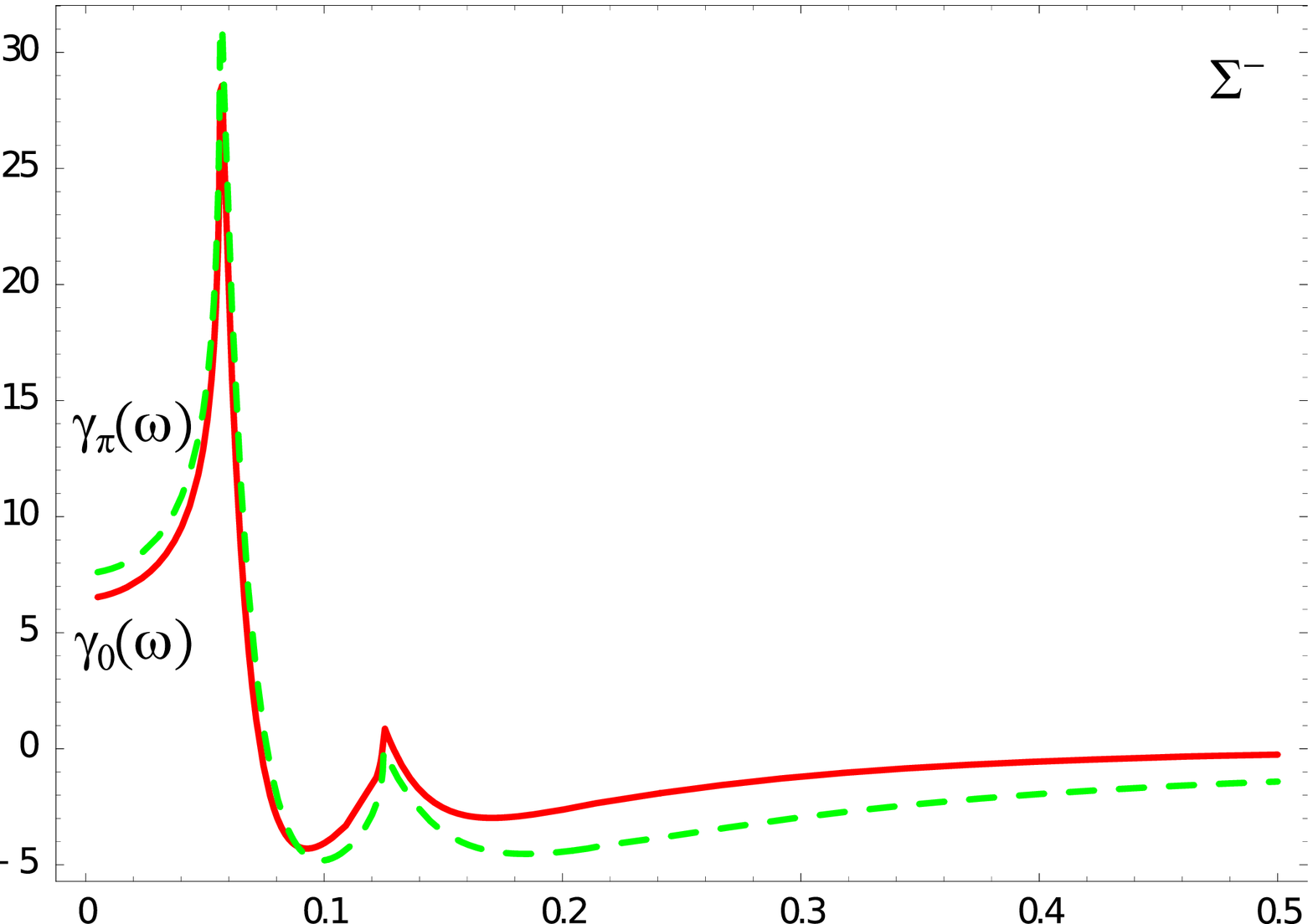}\tabularnewline
\includegraphics[scale=0.2]{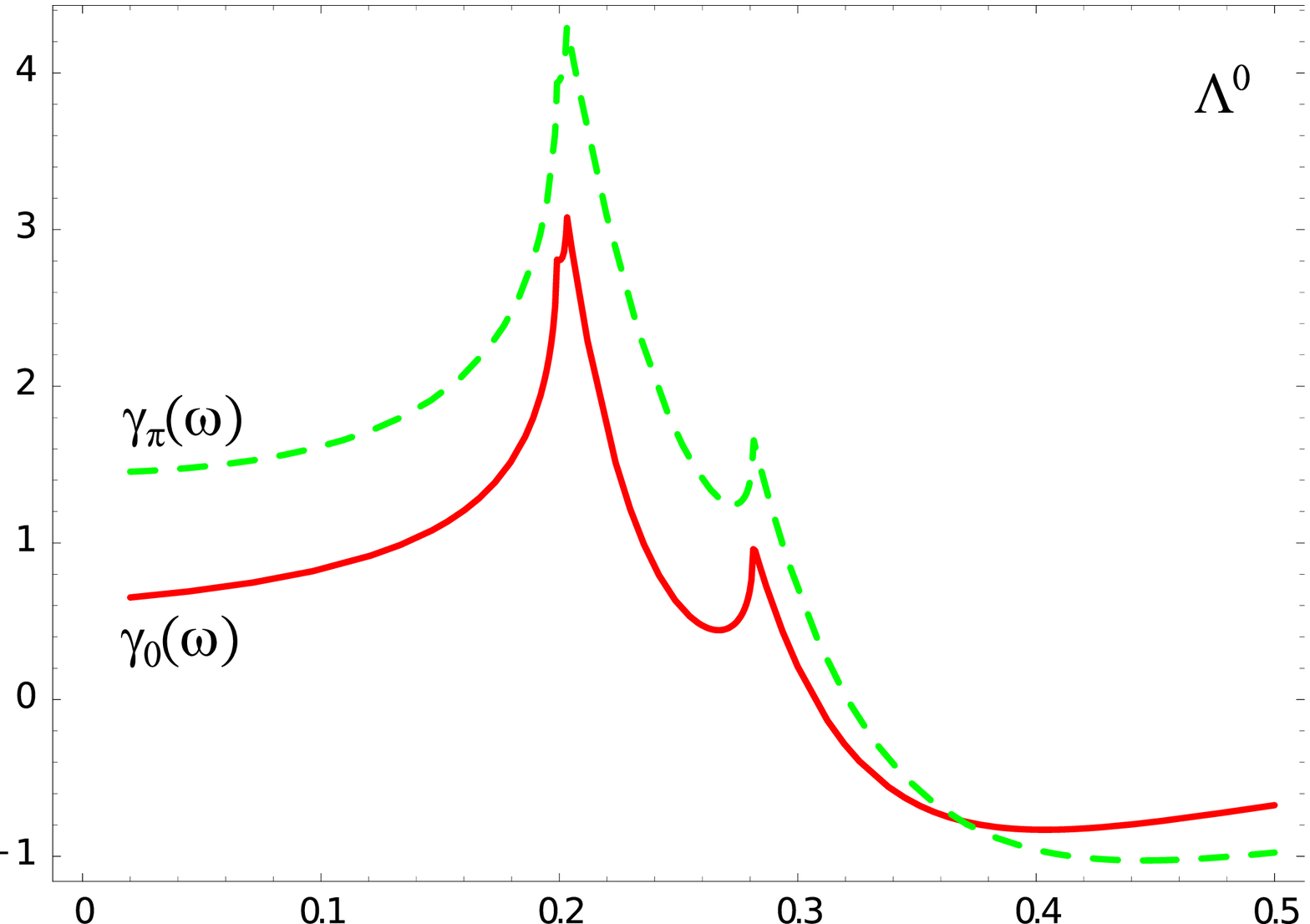} & \includegraphics[scale=0.2]{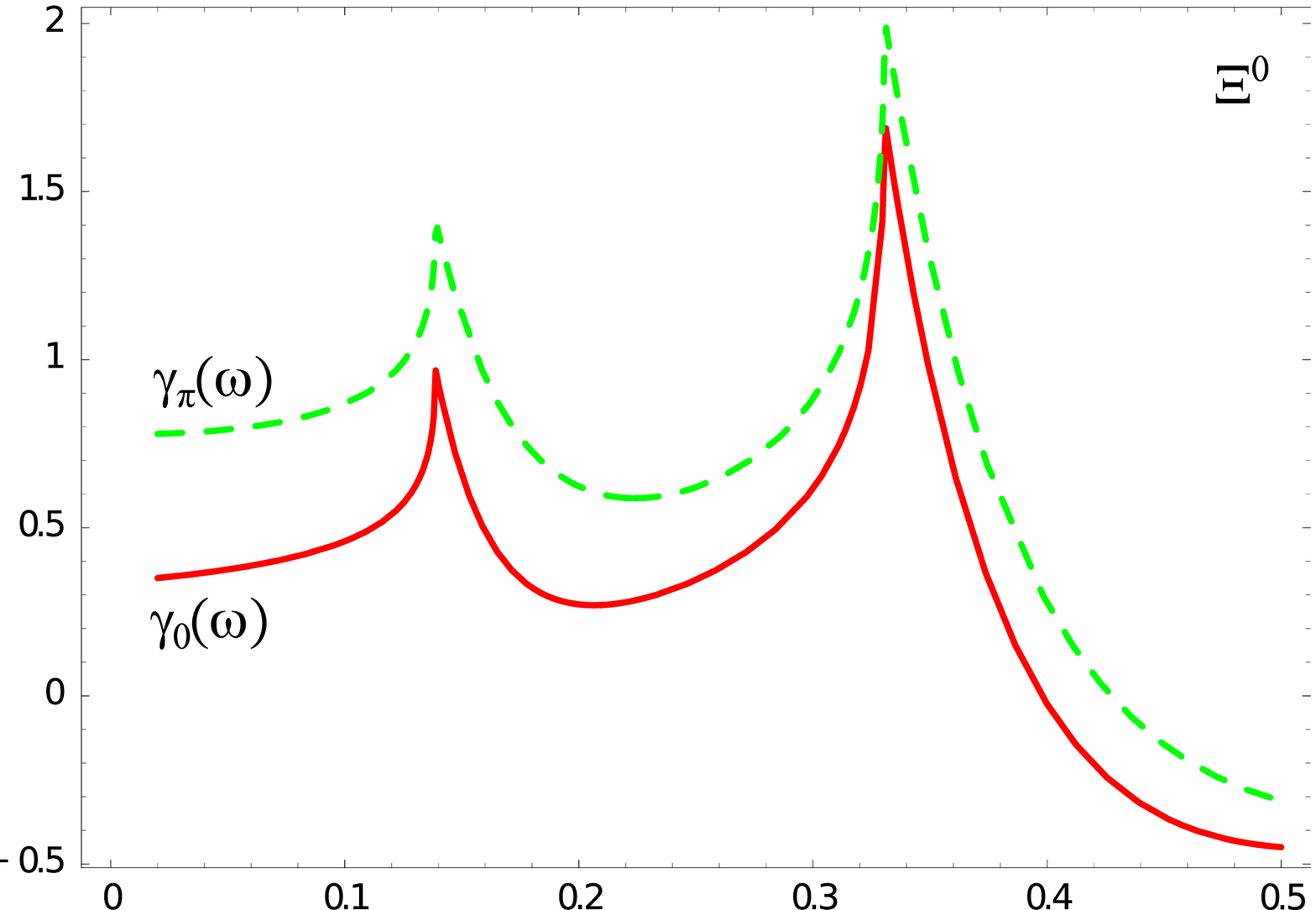} & \includegraphics[scale=0.2]{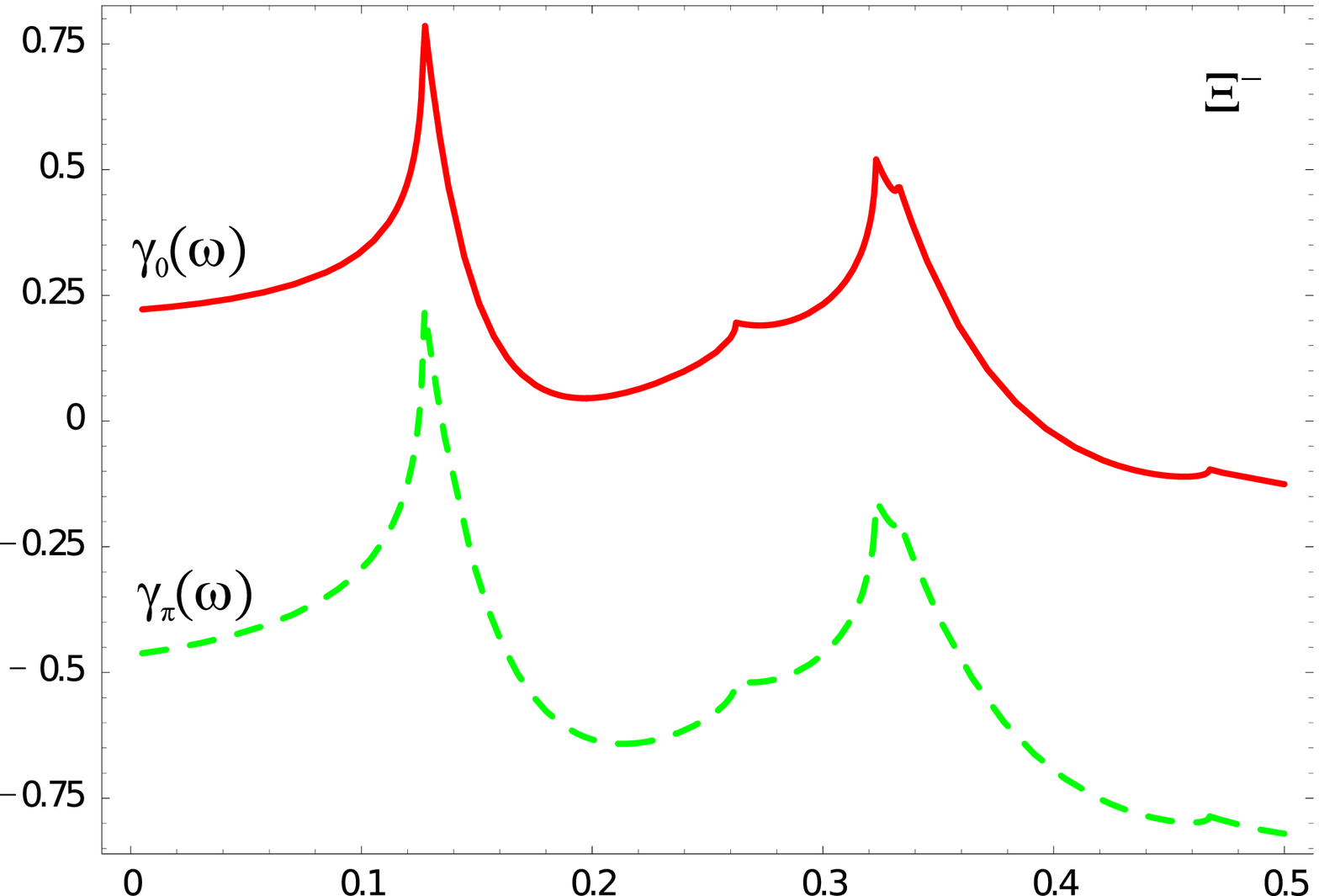}\tabularnewline
\end{tabular}
\par\end{centering}

\caption{Forward $(\gamma_{0})$ and backward $(\gamma_{\pi})$ spin-dependent
dynamical polarizabilities of hyperons in units of $10^{-4}\,(fm^{4})$.
Red solid line shows the forward spin-dependent polarizability and green
dashed line corresponds to the backward spin-dependent polarizability.}
\label{ff3}
\end{figure}

For all polarizabilities listed in Fig.(\ref{ff2}), the electric  polarizabilities have
very similar resonant-type behavior near the meson-production thresholds
and the magnetic polarizabilities for all hyperons have negative low energy
(static) values. Once again it is important to include both pole and
loop resonance contributions for a complete analysis. In Fig.(\ref{ff3}),
we present the first results on the forward and backward
spin-dependent dynamical polarizabilities for the hyperons.

As one can see from Fig.(\ref{ff3}), for all hyperons, the spin-dependent backward polarizability dominates the forward polarizability
on the absolute scale. Simultaneously, they all exhibit almost static
behavior in the very low energy region of the Compton scattering. 
For the all dynamical polarizabilities of SU(3) octet of baryons, we
find that their values are strongly governed by the excitation mechanism
reflected in the meson production peaks. Hence the study of these
polarizabilities directly probes the internal degrees of freedom which
govern the structure of baryons at low energy. 
In this work, we have calculated the electric, magnetic and spin-dependent
dynamical polarizabilities of the SU(3) octet of baryons using ChPTh
implemented in CHM. We found that predictions of the chiral theory
derived from our calculations (up to one-loop order and not including
resonances in loop calculations) are somewhat consistent with the
experimental results. The calculations of the dynamical polarizabilities
with baryon resonances in the loops is our current goal. It is also evident that further
experimental work is needed, especially for the hyperon
polarizabilities.


\begin{thebibliography}{99}
\bibitem{PDG}J. Beringer et al. (Particle Data Group), Phys. Rev.
D86, 010001 (2012).

\bibitem{Pasquini}B. Pasquini, D. Drechsel, and M. Vanderhaeghen,
Eur. Phys. J. Special Topics 198, 269\textendash{}285 (2011).

\bibitem{Babusci}D. Babusci, G. Giordano, A.I. L'vov , G. Matone,
A.M. Nathan, Phys.Rev. C58 (1998) 1013-1041

\bibitem{Griesshammer}R. P. Hildebrandt, H. W. Griesshammer, T. R.
Hemmert, B. Pasquini, Eur. Phys. J A 20, 293-315 (2004).

\bibitem{CHM}A. Aleksejevs, M. Butler, J.Phys. G37 (2010) 035002

\bibitem{SSE}T. R. Hemmert, B. R. Holstein, J. Kambor, Phys. Rev.
D 55, 5598 (1997).

\bibitem{multi-1}V.I. Ritus, ZhETP 32, 1536 (1957) {[}Sov. Phys.
JETP 5, 1249 (1957){]}. 



\bibitem{LenPas}Lensky \& Pascalutsa Eur.Phys.J.C65:195-209, (2010).

\bibitem{OP3}T. R. Hemmert, B. R. Holstein, J. Kambor and G. Knochlein,
Phys. Rev. D 57, (1998) 5746. 

\bibitem{OP4}K. B. Vijaya Kumar, J. A. McGovern and M. C. Birse,
Phys. Lett. B 479, (2000) 167. 

\bibitem{SSE2}G. C. Gellas, T. R. Hemmert and U. G. Meissner, Phys.
Rev. Lett. 86, (2001) 3205. 

\bibitem{HYPDR}D. Drechsel, B. Pasquini and M. Vanderhaeghen, Phys.
Rept. 378, (2003) 99. 

\bibitem{Meissner}V. Bernard, N. Kaiser, J. Kambor and U. Meissner,
Phys. Rev. D 46, 7 (1992).

\bibitem{Kumar}K.B. Kumar, A. Faessler, T. Gutsche, B. Holstein,
V. Lyubovitskij, Phys.Rev. D84, 076007 (2011). 

\bibitem{AB}A. Aleksejevs and S. Barkanova, J.Phys. G38 (2011) 035004

\end{thebibliography}
\end{document}